\documentclass[12pt]{article} 
\usepackage{epsfig}

\renewcommand{\thefootnote}{\fnsymbol{footnote}}

\makeatletter
\def\slash#1{{\mathpalette\c@ncel{#1}}} 
\makeatother
\newcommand{\ub}{\bar u}
\newcommand{\gp}{\gamma_5}
\newcommand{\pro}{{\cal P}}
\newcommand{\gu}{\gamma_\alpha}
\newcommand{\go}{\gamma^\alpha}
\newcommand{\slp}{\slash{p}}

\newcommand{\quark}{\langle \bar q q \rangle}
\newcommand{\ungefaehr}{\,\raisebox{-1ex}{$\stackrel{\textstyle <}{\sim}$}\,}
\def\beq{\begin{equation}}
\def\bea{\begin{eqnarray}}
\def\eeq{\end{equation}}
\def\eea{\end{eqnarray}}
\def\intk{\int{\rd k \over(2\pi)^D}}
\def\al{\alpha}
\def\be{\beta}
\def\bx{\bar u}

\def\rd{{\rm d}^D}
\def\nn{\nonumber}

\begin{document}

\begin{titlepage}
\begin{flushright}
\begin{tabular}{l}
IPPP/03/47 \\
DCPT/03/94\\
August 2003
\end{tabular}
\end{flushright}
\vskip0.5cm
\begin{center}
   {\Large\bf QCD Sum Rules on the Light-Cone,
    Factorisation \\[10pt]
 and SCET}
    \vskip1.3cm {\sc
Patricia Ball\footnote{Patricia.Ball@durham.ac.uk}}
  \vskip0.2cm
	IPPP, Department of Physics, 
University of Durham, Durham DH1 3LE, UK\\ 
  \vskip2cm


\vskip4cm

{\large\bf Abstract:\\[10pt]} \parbox[t]{\textwidth}{ 
The accurate calculation of weak heavy-to-light form factors is crucial 
for the determination of CKM matrix elements from experimental data on
B decays. In SCET, the soft-collinear effective theory, these form
factors can, in the heavy quark limit,
be split into a factorisable part that is calculable in
perturbation theory, and a nonfactorisable part that observes certain
spin-symmetries. I discuss the relation between the SCET factorisation
formulas and the heavy quark limit of the corresponding QCD sum rules on
the light-cone. I also analyse the numerical size of 
factorisable and nonfactorisable parts and corrections suppressed by
powers of the b quark mass.}
 \vfill 
\end{center}
\end{titlepage}

\renewcommand{\thefootnote}{\arabic{footnote}}
\setcounter{footnote}{0}

\section{Introduction and a Bit of History}

The recent interest in B decay form factors into light mesons 
is driven, on the one hand, by the
ever increasing accuracy of experimental results on semileptonic and
rare decays obtained at BaBar and Belle, which calls for
a match in theory, and on the other hand by
recent theoretical developments that aim at a rigorous derivation of
factorisation formulae for form factors in terms of perturbatively
calculable scattering and kernels and  meson distribution
amplitudes on the light-cone. 
Most presently available numerical predictions for form
factors
come from either lattice calculations \cite{lattice} or QCD
sum rules on the light-cone \cite{LCSRFFs}. The
key characteristics of both methods is probably the fact that they give 
predictions for form factors at physical meson masses without any
explicit reference to heavy quark expansion. 
A second, rather different approach is based on the
interpretation of $m_b$ as a large scale which warrants an
expansion in $1/m_b$ and makes it possible to draw
on the power of factorisation theorems and (spin-) symmetry relations 
between form factors for different processes. The first attempt in
this direction
is probably Ref.~\cite{Brodsky}, where $B\to\pi$ was treated in a (as
it turned out) na\"\i\/ve generalisation of the Brodsky-Lepage formalism
for describing exclusive hard perturbative QCD processes
\cite{exclusive}. The factorisation formula derived in \cite{Brodsky}
suffers from a divergence in the convolution of the perturbatively
calculated amplitude with the leading nonperturbative
contributions of the B and $\pi$ mesons; divergences of
this type have been termed {\em soft} or {\em endpoint}
singularities. It was later
argued in Ref.~\cite{CZ} that for decays of heavy mesons the
perturbative hard-gluon exchange process whose dominance is assumed in
the Brodsky-Lepage formalism is actually of the same order in a $1/m_b$
expansion as the so-called soft Feynman mechanism, where not all
partons in a meson participate in the hard subprocess -- in particular
configurations where the spectator quark stays soft, but the light
quark produced in the weak decay has large energy are not suppressed
by powers of $1/m_b$
\begin{figure}[b]
$$\epsfxsize=0.6\textwidth\epsffile{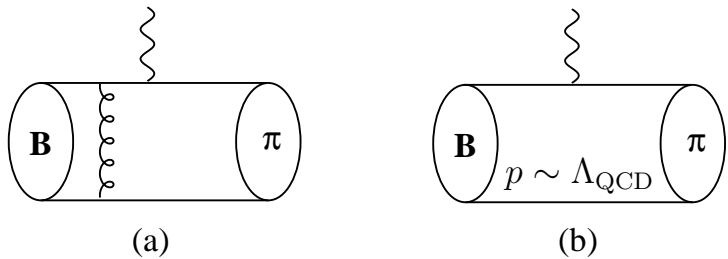}$$
\caption[]{Hard-scattering contribution to $B\to\pi$ form factors
  (a) vs.\ soft contribution (b).}\label{fig:1}
\end{figure}
with respect to the hard-gluon exchange configuration, cf.\ Fig.~\ref{fig:1}.
It was argued, though, that, as a highly asymmetric
edge-of-phase-space configuration, soft
contributions may be suppressed by Sudakov logarithms, but
it was also argued that the logarithmic suppression may not be
effective at the  moderate scale $m_b\sim 5\,$GeV.

The results of Ref.~\cite{CZ} had no immediate impact and 
later papers, e.g.\ Ref.~\cite{hard},
continued to assume dominance of the hard mechanism and concentrated on the
treatment of large logarithms, neglecting power-suppressed effects. The
numerical results for form factors were very small and  
at variance with those from quark models,
Ref.~\cite{BSW}, and lattice calculations and also hardly compatible with 
experimental data on semileptonic B decays. At this point QCD sum
rules on the light-cone (LCSRs) entered the game, 
a generalisation of the original QCD sum rule
method of Shifman, Vainshtain and Zakharov \cite{SVZ} to processes with large
momentum transfer. LCSRs are 
particularly apt to describe {\em both} soft {\em and}
hard contributions to B decays into light mesons \cite{LCSRFFs} --
they can be characterised as combining (field-theoretical) rigour with
a justifiable degree of model-dependence, which makes it possible to
obtain numerical predictions at the physical (b quark) mass scale
without recourse to heavy quark expansion, but at the same time also
limits the accessible accuracy upon inclusion of higher
corrections. The next step forward was stimulated by the success of 
heavy quark effective theory and its spin-symmetry
relations for heavy-to-heavy form factors. Similar relations were
found to apply to heavy-to-light decays, in the combined limits
$m_b\to\infty$ and $E_\pi\to\infty$ and were formulated 
in the framework of LEET (large
energy effective theory) \cite{LEET}, an effective field theory based on the
combination of both heavy quark and large energy limit for light
quarks. The LEET symmetry relations were in
excellent numerical agreement with the results obtained from QCD sum
rules on the light-cone, but neglected the (symmetry-breaking) effects
from hard gluon exchange. The latter were first investigated on a
diagrammatical level in Ref.~\cite{BF} and later on formulated in the
language of effective field theories in Ref.~\cite{SCET}; the
resulting field theory has become known as SCET -- soft-collinear
effective theory. SCET is basically a further development of LEET that
cures its major deficiency, its failure to reproduce the
infrared structure of QCD, which makes it for instance impossible to
properly match LEET onto QCD and results in the (re-)appearance of endpoint
singularities in the hard-gluon exchange contributions to form
factors. SCET is actually not one, but two effective theories with
parametrically different cutoff scales $O(m_b)$ and
$O(\sqrt{m_b\Lambda_{\mbox{\scriptsize QCD}}})$,
respectively, and different degrees of freedom:
(ultra-)soft and collinear quark
and gluon fields in SCET$_{\rm I}$ and (ultra-)soft fields in
SCET$_{\rm II}$. SCET provides a powerful framework to derive and
study factorisation theorems in QCD, allowing one to resum
 large single and double logarithms in the respective 
cutoff scales:
replacing the one-step matching QCD$\leftrightarrow$LEET
by the two-step matching
QCD$\leftrightarrow$SCET$_{\rm I}\leftrightarrow$SCET$_{\rm II}$
removes endpoint singularities (at least to leading order in the
expansion) and allows one to use RG-techniques to resum
(both single and double) logarithms in the parametrically large ratio
of cutoff scales. On the other hand, like with all effective field
theories, a systematic expansion in inverse powers of the
cutoff is possible, but becomes rather
involved, cf.~\cite{sup,sup2}. 
As far as phenomenology is concerned, the question if the correct treatment of 
logarithmic terms $\sim\ln (
m_b/\sqrt{m_b \Lambda_{\mbox{\scriptsize QCD}}})$ at the price of a
considerable complication in the treatment of power-suppressed terms
is justified is a legitimate one -- and yet unanswered. 
It is indeed this question which was one of the main motivations for
writing this paper. 

 One of the major achievements of SCET is the elegant ease with which
the factorisation of amplitudes or form factors, i.e.\ the decoupling
of soft and collinear degrees of freedom, can be formulated as consequence
of simple field transformations of operators in the effective
Lagrangian which decouple (ultra-)soft from collinear fields.
 Factorisation proofs at the Lagrangian level
 are automatically valid to all orders in perturbation theory and as
       such intriguing from a field-theoretical point of view,
although the practitioner will probably be more grateful for explicit LO or NLO
    expressions. Much of this paper will deal will the discussion
of the factorisation formula for one $B\to\pi$ form factor derived in
Ref.~\cite{fuck} and its detailed comparison with the corresponding QCD
sum rule on the light-cone. With SCET's emphasis on factorisation, it
    is not surprising that the categories of ``hard'' and ``soft''
			 contributions which
	   are appropriate and well-defined  in LCSRs, are
     less convenient in a SCET context and are replaced by
 ``factorisable'' and ``nonfactorisable''. But also these categories,
by themselves, are still subject to a certain arbitrariness as for the
definition of the factorisation scheme. Nevertheless, to leading order
  in the heavy quark expansion, SCET spin-symmetry is effective and
   provides a natural factorisation scheme, with symmetry-breaking
corrections induced by short-distance hard-gluon exchange and calculable in
	perturbation theory \cite{BF}. As it was shown in
Ref.~\cite{sup}, and as I shall explicitly demonstrate in this paper,
       SCET spin-symmetry is broken by nonfactorisable terms at
higher order in the heavy quark expansion, which limits the
usefulness of symmetry-relations to relate experimentally measured
form factors to unknown ones.

Let me summarise the questions to be addressed in this paper:
\begin{itemize}
\item What is the relation between factorisation formulas and
  expressions for form factors from QCD sum rules on the light-cone?
\item How large are factorisable contributions to form factors as
  compared to nonfactorisable terms?
\item  How large are power-suppressed corrections to form factor relations?
\end{itemize}
The outline of the paper is as follows: in Sec.~2 I define the
relevant form factors and give a short review of the basics of QCD sum
	   rules on the light-cone as well as of $\pi$ and B
	   distribution amplitudes on the light-cone.
 In Sec.~3 I review the SCET
	   factorised predictions for $B\to\pi$ form factors and
	   relate them to LCSR predictions. In Sec.~4 I discuss
	   the numerics of factorised expressions and
	   power-suppressed terms, and in Sec.~5 I present a summary
	   and conclusions.

\section{Definitions and Framework}

$B\to\pi$ decays can be described by three form factors, $f_+$, $f_0$
and $f_T$, which are defined by ($q=p_B-p$, $q^2 = m_B^2-2 m_B E_\pi$)
\begin{eqnarray}
\lefteqn{\langle \pi(p) | \bar u\gamma_\mu b | B(p_B)\rangle = f_+(q^2) \left\{
(p_B+p)_\mu - \frac{m_B^2-m_\pi^2}{q^2} \, q_\mu \right\} +
\frac{m_B^2-m_\pi^2}{q^2} \, f_0(q^2)\, q_\mu,}\hspace*{5.3cm}\\
\langle \pi(p) | \bar d \sigma_{\mu\nu} q^\nu (1+\gamma_5) b | B(p_B)\rangle
& \equiv &  \langle \pi(p) | \bar d \sigma_{\mu\nu} q^\nu b |
B(p_B)\rangle\nonumber\\
& = & i\left\{ (p_B+p)_\mu q^2 - q_\mu (m_B^2-m_\pi^2)\right\} \,
  \frac{f_T(q^2)}{m_B+m_\pi}. 
\end{eqnarray}
In the context of LCSRs, it is convenient to introduce one more form
factor:
\begin{equation}\label{eq:deffmi}
f_-(q^2) = \frac{m_B^2-q^2}{q^2}\, (f_0(q^2) - f_+(q^2)).
\end{equation}
Note that $f_+(0)=f_0(0)$. 
For large energies of the final state $\pi$, i.e.\ $E_\pi \sim
O(m_b/2)$, the form factors can be calculated from QCD sum rules on the
light-cone. This method combines standard QCD sum rule techniques with
the information on light-cone hadron distribution amplitudes (DAs)
familiar from the theory of exclusive processes. 
The key idea is to
consider a correlation function of the weak current and a current with
the quantum-numbers of the B meson, sandwiched between the vacuum and
the $\pi$. For large (negative) virtualities of these currents, the
correlation function is, in coordinate-space, dominated by light-like distances
 and can be expanded around the
light-cone. In contrast to the short-distance expansion
employed in conventional QCD sum rules \`a la SVZ \cite{SVZ}, where
nonperturbative effects are encoded in vacuum expectation values
of local operators with
vacuum quantum numbers, the condensates, LCSRs
rely on the factorisation of the underlying correlation function into
genuinely nonperturbative and universal hadron DAs
$\phi$. The DAs are convoluted with process-dependent amplitudes $T_H$,
which are the analogues to Wilson-coefficients in
short-distance expansion and can be
calculated in perturbation theory, schematically
\begin{equation}\label{eq:3}
\mbox{correlation function~}\sim \sum_n T_H^{(n)}\otimes \phi^{(n)}.
\end{equation}
The sum runs over contributions with increasing twist, labelled by
$n$, which are suppressed by
increasing powers of, roughly speaking, the virtualities of the
involved currents.
The same correlation function can, on the other hand, be written as a
dispersion-relation, in the virtuality of the current coupling to the
B meson. Equating dispersion-representation and
light-cone expansion, and separating the B meson contribution from
that of higher one- and multi-particle states, one obtains a relation
(QCD sum rule) for the form factor.

For $B\to\pi$ form factors the relevant correlation function is
\begin{eqnarray}\label{eq5}
i\int d^4y e^{-ip_By} \langle\pi(p)|TJ_\mu(0) j_B^\dagger(y)|0\rangle
= \Pi_+(p_B+p)_\mu + \Pi_-(p_B-p)_\mu
\end{eqnarray}
 with $j_B = \bar d i\gamma_5 b$ and $J_\mu = \bar u\gamma_\mu b$ for
 $f_{+,0}$ and $J_\mu = \bar d \sigma_{\mu\nu} q^\nu b$ for $f_T$. 
LCSRs for all three form factors are available at $O(\alpha_s)$ 
accuracy for the twist-2 and tree-level accuracy for twist-3 and 4
 contributions \cite{NLOK,NLOB}. Radiative corrections to the
 2-particle twist-3 contributions to $f_+$ have been calculated in
 Ref.~\cite{roman}. For this paper, I have also calculated the
 corresponding corrections for $f_-$ and $f_T$. The
correlation function $\Pi_+$, calculated for unphysical
$p_B^2$, can be written as dispersion-relation over its physical cut. Singling
out the contribution of the B-meson, one has, for the vector current
 $J_\mu = \bar u \gamma_\mu b$,
\begin{equation}\label{eq:corr}
\Pi_+ =  f_+(q^2) \, \frac{m_B^2f_B}{m_B^2-p_B^2}
+ \mbox{\rm higher poles and cuts},
\end{equation}
where $f_B$ is the leptonic decay constant of the B-meson,
$f_Bm_B^2=m_b\langle B| \bar b i\gamma_5 d|0\rangle$.
In the framework of LCSRs one does not use (\ref{eq:corr}) as it stands,
but performs a  Borel-trans\-for\-ma\-tion,
\begin{equation}\label{eq:9}
\hat{B}\,\frac{1}{t-p_B^2} = \frac{1}{M^2} \exp(-t/M^2),
\end{equation}
with the Borel-parameter $M^2$; this transformation enhances the
ground-state B-meson contribution to the dispersion-representation of $\Pi_+$
and suppresses contributions of higher twist to its light-cone expansion.
The next step is to invoke quark-hadron
duality to approximate the contributions of hadrons other than the
ground-state B-meson by the imaginary part of the light-cone
expansion of $\Pi_+$, so that
\begin{eqnarray}
\hat{B}{\Pi_+^{\rm LC}} & = &
\frac{1}{M^2}\, m_B^2f_B \,f_+(q^2)\,e^{-m_B^2/M^2} +
\frac{1}{M^2}\, \frac{1}{\pi}\int_{s_0}^\infty \!\! dt \, {\rm
Im}{\Pi^{\rm LC}_+}(t) \, \exp(-t/M^2)\\
{\rm and}\quad \hat{B}_{\rm sub}\Pi_+^{\rm LC} & = & \frac{1}{M^2}\,
m_B^2f_B \,f_+(q^2)\,e^{-m_B^2/M^2}.\label{eq:SR}
\end{eqnarray}
Eq.~(\ref{eq:SR}) is the LCSR for $f_+$. 
$s_0$ is the so-called continuum
threshold, which separates the ground-state from the continuum
contribution. At tree-level, the continuum-subtraction in
(\ref{eq:SR}) introduces a lower limit of integration, $u\geq
(m_b^2-q^2)/(s_0-q^2)\equiv u_0$, in (\ref{eq:3}), which behaves as
$1-\Lambda_{\rm QCD}/m_b$ for
large $m_b$ and thus corresponds to the dynamical
configuration of the Feynman-mechanism, as it cuts off low momenta of
the u quark created at the weak vertex. This is how soft contributions
enter LCSRs. At $O(\alpha_s)$, there are
also contributions with no cut in the integration over $u$, which
correspond to hard-gluon exchange contributions. 
As with standard QCD sum rules, the use of quark-hadron 
duality above $s_0$ and the
choice of $s_0$ itself introduce a certain model-dependence (or
systematic error) in the final result for the form-factor.

Let me shortly comment on the expansion parameter of LCSRs. It has
been claimed that the expansion goes in powers of $1/m_b$, e.g.\
\cite{Lange2}. The situation is not that simple, however. In the next
section we shall see that twist-3 DAs contribute at the same order in
$1/m_b$ as twist-2 ones -- this observation is not really new and has
first been made in Ref.~\cite{simma}. It actually turns out that the
coefficients of twist-3 and 4 amplitudes, as far as they are known, do
not show any suppression by powers of $m_b$ at all, but that it is the
shape of distribution amplitudes as predicted by conformal expansion
that entails an effective suppression in inverse powers of $m_b$ upon
(soft) convolution over $u$. This suppression becomes effective only if
an {\em explicit} $1/m_b$ expansion is performed (as we shall do in
the next section). In addition, tree-level power-counting may be upset by
radiative corrections: it turns out that the tree-level 3-particle
twist-3 contribution is suppressed by $1/m_b$ with respect to the twist-2
contribution, but this suppression is not removed by $O(\alpha_s)$
corrections. In general, for $\pi$ DAs, we expect that higher twist
DAs are dominated by {\it gluonic} DAs, and that their matrix elements
are suppressed numerically (not parametrically) by powers of $\alpha_s$.

To twist-3 accuracy, the DAs of the $\pi$ are defined as ($\ub = 1-u$)
\begin{eqnarray}
\langle \pi | \bar u(0)\gamma_\mu \gamma_5[0,x] u(x)|0\rangle & = & -ip_\mu
f_\pi \int\!\! du\, e^{i\bar u px} \phi_\pi(u),\\
\langle \pi | \bar u(0)i\gamma_5[0,x]d(x)|0\rangle & = & \mu_\pi^2 
\int \!\! du\, e^{i\bar u px} \phi_p(u),\\
\langle \pi | \bar u(0)\sigma_{\alpha\beta} i\gamma_5[0,x] d(x)|0\rangle 
& = & \frac{1}{6}\,\mu_\pi^2 (p_\alpha x_\beta - x_\alpha
p_\beta) \int \!\! du\, e^{i\bar u px} \phi_\sigma(u),\nonumber\\
\lefteqn{\langle \pi | \bar u(0)\sigma_{\alpha\beta}\gamma_5[0,vx] 
gG_{\mu\nu}(vx)[vx,x] d(x) | 0 \rangle =}\hspace*{6cm}\nonumber\\ 
\lefteqn{ =  i\mu_\pi^2 \left[(p_\mu p_\alpha
g_{\nu\beta} - (\mu\leftrightarrow\nu)) - (p_\mu p_\beta
g_{\nu\alpha} - (\mu\leftrightarrow\nu))\right] \int{\cal
D}\alpha \, {\cal T}(\underline{\alpha})
e^{ipx(\alpha_2+v\alpha_3)}.}\hspace*{6cm}\label{eq:Tgluon}
\end{eqnarray}
Here $[x,y]$ is the path-ordered gauge-factor
$$
[x,y] = {\rm P}\,\exp\left[ig\int_0^1 dt (x-y)_\mu A^\mu(tx + (1-t)
  y)\right],
$$
$\mu_\pi^2 = f_\pi m_\pi^2/(m_u+m_d) = -2\quark/f_\pi$ 
and ${\cal D}\alpha = d\alpha_1 d\alpha_2 d\alpha_3 \delta(1-\alpha_1-
\alpha_2-\alpha_3)$. $\phi_\pi$ is a twist-2 DA, $\phi_\sigma$,
$\phi_p$, ${\cal T}$ are, loosely speaking, twist-3, although $\phi_\sigma$
and $\phi_p$ also contain admixtures of twist-2 contributions. The DAs
can be expressed as a partial wave expansion in conformal spin
(conformal expansion); more details and explicit expressions can be
found in Ref.~\cite{update}.
For the present study, we mostly need the so-called
asymptotic DAs, the leading term in the partial wave expansion:
\begin{equation}\label{asymptotic}
\phi_\pi(u) = \phi_\sigma(u) = 6 u (1-u),\quad \phi_p(u) = 1, \quad
    {\cal T}(\underline{\alpha}) = 360\eta_3(\mu) \alpha_1 \alpha_2 \alpha_3^2,
\end{equation}
where the hadronic matrix element 
$\eta_3(\mu)$ is defined by the local limit of (\ref{eq:Tgluon}).
To higher order in the expansion, also the 2-particle
DAs become scale-dependent and 
observe evolution equations. We will need in particular the evolution 
equation of $\phi_\pi'(1)$, first derived in \cite{NLOB} ($a_s =
C_F\alpha_s/(4\pi)$):
\begin{equation}\label{eq:evolvephipi}
\mu\,\frac{d}{d\mu}\,\phi'_\pi(1,\mu) = -4a_s \left\{
\int_0^1 du \left[ \frac{\phi_\pi(u) + \ub \phi_\pi'(1)}{\ub^2} +
  \frac{\phi_\pi(u)}{\ub} \right] - \frac{1}{2}\,\phi_\pi'(1)\right\}.
\end{equation}
Note that the three twist-3 DAs are not independent of each other.
Solving the recursion relation for the moments of e.g.\ $\phi_p(u)$,
given in \cite{update}, I find ($\xi = 2u-1$):
\begin{eqnarray}
\frac{d}{du}\,\phi_p(u) & = & \frac{1}{1-\xi^2}\,\left(\xi\,
\frac{d^2}{du^2}\,\phi_{{\cal T}_1}(u) - 2\,\frac{d}{du}\,\phi_{{\cal
    T}_1}(u) + \,\frac{d^2}{du^2}\,\phi_{{\cal T}_2}(u)\right),\label{eq:DGE}\\
\mbox{with~}\phi_{{\cal T}_1}(u) & = & \int_0^u d\alpha_1
\int_0^{\ub} d\alpha_2\,\frac{2}{\alpha_3}\,{\cal
  T}(\underline{\alpha}),\quad \phi_{{\cal T}_2}(u) =  \int_0^u d\alpha_1
\int_0^{\ub} d\alpha_2\,\frac{2(\alpha_1-\alpha_2-\xi)}{\alpha_3^2}\,{\cal
  T}(\underline{\alpha}).\nonumber
\end{eqnarray}
The initial condition for the differential equation (\ref{eq:DGE}) is
given by the recursion relation for the second moment of $\phi_p$,
given in \cite{update}. Solving
(\ref{eq:DGE}), I find:
\begin{equation}
\phi_p(u,\mu) = 1 + 12\eta_3(\mu) + \frac{1}{2}\,\int_0^1
dv\,(2v-1)^3\,\frac{d}{dv}\, \phi_p(v,\mu) +
\int_0^u dv\,\frac{d}{dv}\,\phi_p(v,\mu),
\end{equation}
which expresses $\phi_p$ uniquely in terms of the 3-particle DA
${\cal T}$. A similar relation can be derived for $\phi_\sigma$. The
evolution equation for ${\cal T}$ is not known.

For completeness, let me also introduce the leading-twist 
B meson DA which features
in SCET factorised expressions (but not in LCSRs). It can be defined
as\footnote{To be precise, there are {\em two} B DAs, but only one of
  them enters the leading-order 
SCET factorised expressions for $B\to\pi$ form factors.}
\cite{Lange1}
\begin{equation}\label{eq:BDA}
\langle 0 | \bar q_s(x) [x,0]_s \slash{n}\Gamma h(0)|B(v)\rangle =
-i\, \frac{F(\mu)}{2}\,{\rm
  Tr}(\slash{n}\Gamma\,\frac{1+\slash{v}}{2}\,\gamma_5)
\int_0^\infty d\omega\,e^{-i\omega (v\cdot x)} \phi_+^B(\omega,\mu),
\end{equation}
where $n$ is a light-like vector parallel to $x$ and $F(\mu)$ is the
matrix element corresponding to the asymptotic value of $\sqrt{m_B}
f_B$ in heavy quark effective theory (HQET); $v$ is the four-velocity
of the B meson. The index $s$ denotes the soft modes of
(quark and gluon) fields in SCET and $h$ is the heavy quark field in HQET. 
The normalisation of $\phi_+^B$ is chosen such that
$$
\int_0^\infty d\omega \phi_+^B(\omega, \mu) = 1.
$$
The scale-dependence of $\mu$ has been discussed in \cite{Lange1},
with the result that running from an initial scale $\mu_0$ to $\mu$
induces a radiative tail of the DA that falls off slower than
$1/\omega$ at large $\omega$, which renders positive moments of the DA
ill-defined, independent of its behaviour at $\mu_0$. In SCET
factorised expressions, however, $\phi_+^B(\omega)$ always enters in
the combination $\phi_+^B(\omega)/\omega$, modulo logs, i.e.\ as
$$
\int_0^\infty d\omega\,\frac{\phi_+^B(\omega)}{\omega}\,
\ln^n\,\frac{\omega}{\mu},
$$
which is well-defined at all scales. For $n=0$, a
simple evolution equation emerges.
Defining
$$\frac{1}{\lambda_B(\mu)} = \int_0^\infty d\omega\,
\frac{\phi_+^B(\omega,\mu)}{\omega},
$$
I extract the following evolution equation from the expressions given
in Ref.~\cite{Lange1}:
\begin{equation}
\mu\,\frac{d}{d\mu}\,\frac{1}{\lambda_B(\mu)} =
2 a_s\, \frac{1}{\lambda_B(\mu)} + 4a_s \int_0^\infty
d\omega\,\frac{\phi_+^B(\omega,\mu)}{\omega} \,
\ln\,\frac{\omega}{\mu}.
\end{equation}
I have checked that the above evolution equation indeed ensures
$\mu$-independence of the factorised expression
for the $B\to\gamma$ form factor obtained in Ref.~\cite{Bgamma}.

\section{Light-Cone Sum Rules vs.\ SCET -- A Case Study}

In the SCET limit $m_b\to\infty$, $E_\pi\to\infty$, spin-symmetry
ensures that the three $B\to\pi$ form factors can  be expressed in
terms of one single nonperturbative function $\zeta(E_\pi)$ that
includes the soft contributions. Spin-symmetry is broken by two
effects: hard gluonic corrections to the weak vertex, which yield
vertex-specific matching conditions at $\mu=m_b$, whereas the
logarithmic terms $\ln (m_b/\mu)$ are spin-symmetric, and hard spectator
interactions. 

Combining the results of Refs.\ \cite{BF} and \cite{SCET}, the form
factors at $q^2=0$, i.e.\ $E_\pi = m_B/2 = E_{\rm max}$ can be written
as ($a_s = C_F\alpha_s/(4\pi)$):
\begin{eqnarray}
f_+(0) & = & \left[ C_4\left(\frac{m_b}{\mu_F}\right) + \frac{1}{2}\,
  C_5\left(\frac{m_b}{\mu_F}\right) \right] \zeta(E_{\rm max},\mu_F) + a_s
  \,\Delta f_+(\mu_F) + O(1/m_b),\label{eq:fpl}\\
f_-(0) &=& 
-\left[ C_4\left(\frac{m_b}{\mu_F}\right) - \frac{1}{2}\,
  C_5\left(\frac{m_b}{\mu_F}\right) \right] \zeta(E_{\rm max},\mu_F) + a_s
  \,\Delta f_-(\mu_F) + O(1/m_b),\label{eq:fmi}\\
f_T(0,\mu) & = & C_{T}\left(\frac{m_b}{\mu}\right) \left[
C_{11}\left(\frac{m_b}{\mu_F}\right) \zeta(E_{\rm max},\mu_F) + a_s
  \,\Delta f_T(\mu_F)\right] + O(1/m_b).\label{eq:fT}
\end{eqnarray}
The phenomenological interest of factorised expressions like the above
lies in the fact that the nonperturbative quantity $\zeta$ can be
eliminated in  ratios of form factors, which are then
predicted in terms of calculable expressions. The usefulness of such
relations depends on the size of neglected terms; we shall come back
to that question in Sec.~4.
$C_T$ is the standard 
 QCD  Wilson-coefficient for the tensor-current, whereas
 $C_i$ resums large single and double 
logarithms $\sim \ln m_b/\mu_F$ encountered in the matching of
SCET$_{\rm I}$ onto QCD. Explicit resummed 
expressions can be found in \cite{fuck};
expanded to first order in $\alpha_s$, one has
\begin{eqnarray}
C_4\left(\frac{m_b}{\mu_F}\right) & = & \left\{ 1 - a_s \left(
 \frac{\pi^2}{12} + \frac{13}{2}\right)
\right\} \left\{ 1 + a_s\left( -2\ln^2\frac{m_b}{\mu_F} +
5\ln\frac{m_b}{\mu_F} \right)\right\},\nonumber\\
C_5\left(\frac{m_b}{\mu_F}\right) & = & a_s,\nonumber\\
C_{11}\left(\frac{m_b}{\mu_F}\right) & = & \left\{ 1 - a_s 
\left( \frac{\pi^2}{12} + 4\right)
\right\} \left\{ 1 + a_s\left( -2\ln^2\frac{m_b}{\mu_F} +
5\ln\frac{m_b}{\mu_F} \right)\right\},\nonumber\\
C_T\left(\frac{m_b}{\mu}\right) & = & 1 + a_s\ln\frac{m_b^2}{\mu^2}.
\label{wilsons}
\end{eqnarray}
In the notation of Ref.~\cite{BF}, the $\Delta f$ in Eqs.~(\ref{eq:fpl}) to 
(\ref{eq:fT}) denote the contributions from hard-gluon exchange
between the b or u quark (d quark for $f_T$) and the spectator
quark. The corresponding diagrams contain
endpoint singularities, but the authors of Ref.~\cite{BF} argued
that these singularities are spin-symmetric and hence can be
absorbed into $\zeta$. From this it is evident that the 
separation between ``hard'' and ``soft''
contributions is not well-defined eo ipso, but requires the
definition of a factorisation scheme that fixes precisely which terms
are included into $\zeta$ and which into $\Delta f$. In Ref.~\cite{BF},
the factorisation scheme was defined by
$$f_+(q^2) \equiv \xi(E_\pi).
$$
In this scheme also hard-gluon corrections to the weak vertex which
contain double logarithms, but are spin-symmetric, are
absorbed into $\xi$. 
To $O(\alpha_s)$ accuracy, the following relations were obtained:
\begin{eqnarray}
f_-(0) & = & -\xi(E_{\rm max})\left\{ 1 - a_s \right\} +
a_s\, \Delta F\label{eq22}\\
f_T(0,\mu) & = & \frac{m_B+m_\pi}{m_B}\,\xi(E_{\rm max}) \left\{ 1 + a_s
\left( \ln\frac{m_b^2}{\mu^2} + 2\right)\right\} - a_s\, 
\frac{m_B+m_\pi}{m_B}\,\Delta F,\label{eq:blurb}\\
\mbox{with~} \Delta F & = & \frac{8\pi^2}{3}\,\frac{f_B
  f_\pi}{m_B}\,\frac{1}{\lambda_B}
\left\langle\frac{1}{\ub}\right\rangle_\pi.\label{eq:deltaf}
\end{eqnarray}
The results of Ref.~\cite{BF} indicate that the classification of
form factor contributions as ``hard'' or ``soft'' may not be the most
appropriate one for heavy-to-light form factors and should be
replaced by one
distinguishing between spin-symmetric and non-symmetric terms. This is
indeed the pattern emerging in SCET, where the
following  factorisation formula, valid to all orders in $\alpha_s$
and leading order in $1/m_b$, has been derived in Ref.~\cite{fuck}:
\begin{eqnarray}
f_+(q^2) & = &
\left(C_4(E_\pi,m_b/\mu_I)+\frac{E_\pi}{m_B}\,C_5(E_\pi,m_b/\mu_I)
\right)  \zeta (E_\pi,\mu_I)\nonumber\\
&+&{}
\frac{m_Bf_Bf_\pi}{4E_\pi^2} \, \int_0^\infty dk_+\int_0^1 du\, dv\,
\phi_B^+(k_+,\mu_{II})\phi_\pi(u,\mu_{II})\nonumber\\
&\times &
\left( \frac{2E_\pi-m_B}{m_B}\,T_a(\mu_I) \,J_a(u,v,k_+; \mu_I,\mu_{II})
+ \frac{2E_\pi}{m_B}\,T_b(\mu_I) J_b(u,v,k_+;
\mu_I,\mu_{II})\right).\hspace*{0.8cm} \label{eq:SCETfuck}
\end{eqnarray}
Here $\mu_{I(II)}$ is a scale in SCET$_{\rm I(II)}$, i.e.\
$\Lambda_{\rm QCD}\ungefaehr \mu_{II} < 
\sqrt{m_b\Lambda_{\rm QCD}} \ungefaehr \mu_I < m_b.$ $T_i$ are
SCET$_{\rm I}$ Wilson-coefficients of (subleading) currents in the 
SCET$_{\rm I}$ Lagrangian and the jet functions $J_i$ are the
Wilson-coefficients of matching SCET$_{\rm II}$ quark bilinear
operators onto SCET$_{\rm I}$, with, to $O(\alpha_s)$ accuracy \cite{fuck}:
$$T_a J_a = T_b J_b = \frac{4\pi^2}{3}\,a_s\,\frac{\delta(u-v)}{uk_+}\,.$$
The first term in (\ref{eq:SCETfuck}) is nonfactorisable, but 
spin-symmetric,
except for the values of the matching coefficients at $\mu_I =
m_b$. The second term is factorisable and does not contain endpoint
singularities, but breaks spin-symmetry.  One
important consequence of the decoupling of collinear and soft degrees
of freedom in SCET is that the two terms do not mix under a change of
scales. As we shall see below, this suggests that all twist-3 effects
which are of the same order in $1/m_b$ as the twist-2 contributions
can be absorbed into $\zeta$. 
Note that $\zeta_(E_\pi,\mu_F)$  is defined in SCET$_{\rm I}$ and that
also contributions of the second B DA $\phi_B^-$ have been absorbed
into it. The question arises if the fact that spin-symmetric and
nonfactorisable terms coincide is an artifact of the leading order in
the $1/m_b$ expansion. As shown in Ref.~\cite{sup} this is indeed the
case and may limit  the relevance of 
Eq.~(\ref{eq:SCETfuck}) from a phenomenological
point of view, as nonfactorisable (uncalculable)
symmetry-breaking terms at nonleading order in $1/m_b$ cannot be
eliminated any more from form factor relations.

In the remainder of this section, I will
compare the LCSR predictions for $f_{+,-,T}$ with the above
relations and address in particular the following questions:\label{roadmap}
\begin{itemize}
\item do LCSRs reproduce the relations (\ref{eq22}),
(\ref{eq:blurb}) and (\ref{eq:deltaf})?
\item do LCSRs reproduce the logarithmic terms resummed in the
  Wilson-coefficients $C_i$?
\item can (\ref{eq:SCETfuck}) be directly derived from LCSRs (to
  $O(\alpha_s)$)?
\item are there soft symmetry-breaking contributions at nonleading
  order in $1/m_b$?
\end{itemize}

In order to compare light-cone sum rules with the factorisation
formulas predicted by SCET, I have
\begin{itemize}
\item calculated the LCSRs for all three form factors for $B\to\pi$
  transitions in full QCD, at $q^2=0$, i.e.\ $E_\pi = m_B/2$, to
  $O(\alpha_s)$ accuracy for 2-particle twist-2 and 3 contributions; 
\item performed the heavy quark limit in the following way
  \cite{NLOB}: the sum rule parameters $M^2$ and $s_0$ are scaled with the
  heavy quark mass as
\begin{equation}\label{scaling}
M^2 = 2 m_b\tau,\quad s_0-m_b^2 = 2 m_b \omega_0,
\end{equation}
with $\tau$ and $\omega_0\sim 1\,$GeV.
These scaling laws follow from the requirement of the sum rule for
$f_B$ to observe the correct scaling in the heavy quark limit \cite{BBBD}. In
addition, in order to obtain simple expressions, 
I also fix $q^2=0$ ($E_\pi=m_B/2$) and
apply the finite-energy limit $\tau\to\infty$. In this limit,
the B meson corresponds to a heavy quark accompanied by light degrees
of freedom with an energy less or equal $\omega_0$, i.e.\ 
a distribution with a sharp cutoff.
\end{itemize}
The rationale behind this procedure is the observation that
SCET factorisation formulas rely on 
manipulations of the Lagrangian and hence are independent of 
the  realisation of physical states; consequently, 
they must also be realised in LCSRs.

In the combined heavy quark and finite-energy limit 
LCSRs depend on the $\pi$ DAs of leading and
higher twist, $\omega_0$, the only parameter characterising the B
meson, the factorisation scale $\mu$ characteristic for the $\pi$, 
and $m_b$. For $f_+$ and twist-2 accuracy, the
corresponding formula was first obtained in \cite{NLOB}. Radiative
corrections to the twist-3 contributions to $f_+$ have been calculated in
\cite{roman}. Here, we extend these analyses and include also
radiative corrections to twist-2 and twist-3 2-particle DAs for $f_-$
and $f_T$. The corresponding expressions are given in the appendix.
The contributions to $f_+(0)$ read:
\begin{eqnarray}
\lefteqn{
f_B\, m_b\, f_+^{T2}(0)  =  -f_\pi\,\frac{\omega_0^2}{m_b}\, 
\phi_\pi'(1,\mu) \left\{ 1 + a_s \left( 1+\pi^2
-2\ln^2\,\frac{2\omega_0}{m_b} -4\ln\,\frac{2\omega_0}{m_b} + 2
\ln\, \frac{2\omega_0}{\mu}\right)\right\}}\hspace{2cm}\nonumber\\
&&\hspace{-1cm} + 4 a_s\,f_\pi\,\frac{\omega_0^2}{m_b}\left\{\left(
\ln\,\frac{2\omega_0}{ \mu}-1\right) \int_0^1 du\,\frac{\phi_\pi(u)
  +\bar u \phi_\pi'(1)}{\bar u^2} + \ln\,\frac{2\omega_0}{ \mu} \int_0^1
du\,\frac{\phi_\pi(u)}{\bar u}\right\},\hspace*{0.5cm}\label{eq:1}\\
\lefteqn{
f_B\, m_b\, f_+^{T3,p}(0)  =
\mu_\pi^2(\mu)\,\frac{\omega_0}{m_b}\,\phi_p(1,\mu) \left\{ 1 + a_s
\left(\pi^2-4 - 2 \ln^2 \,\frac{2\omega_0}{m_b}
  + 4 \ln\, \frac{2\omega_0}{\mu}\right.
\right.}\hspace{2cm}\nonumber\\
&& \hspace{-1cm}\left.\left.- 4 \ln\, \frac{2\omega_0}{\mu}\,
\ln\,\frac{2\omega_0}{m_b}\right)\right\} + 
2 \mu_\pi^2 \frac{\omega_0}{m_b}\, a_s \left(
4\ln\,\frac{2\omega_0}{\mu} - 3 \right) \int_0^1 du\,\frac{1}{\ub}\,
(\phi_p(u) - \phi_p(1)),\\
\lefteqn{
f_B\, m_b\, f_+^{T3,\sigma}(0)  =
-\mu_\pi^2(\mu)\,\frac{\omega_0}{6m_b}\,\phi_\sigma'(1,\mu) \left\{ 1 + a_s
\left(\pi^2-10 - 2 \ln^2 \,\frac{2\omega_0}{m_b}
-8\ln\,\frac{2\omega_0}{m_b} \right.
\right.}\hspace{2cm}\nonumber\\
&& \hspace{-0.7cm}\left.\left. + 8 \ln\, \frac{2\omega_0}{\mu}
+ 4 \ln\, \frac{2\omega_0}{\mu}\,
\ln\,\frac{2\omega_0}{m_b}\right)\right\} - 
\mu_\pi^2 \frac{\omega_0}{3m_b}\, a_s \int_0^1 du\,\frac{1}{\ub^2}\,
(\phi_\sigma(u) +\ub \phi_\sigma'(1)).
\end{eqnarray}
Let me first discuss $f_+^{T2}(0)$. The first point to notice is that
by virtue of the evolution equation (\ref{eq:evolvephipi}), the
r.h.s.\ of (\ref{eq:1}) is indeed independent of the factorisation
scale $\mu$. Soft and hard contributions can clearly be identified: 
soft terms are those with a highly asymmetric configuration of
the quarks in the $\pi$, i.e.\ the terms in $\phi'_\pi(1)$, whereas
those in $\int_0^1 du$ are hard contributions. Note that both 
contributions combine in such a way that integrands with
$\ub^2$ in the denominator that could give rise to endpoint
singularities are automatically regularised. It is also evident that the
separation between hard and soft contributions is $\mu$-dependent. 

Let us next look at the logarithms in $m_b/\mu$. It is clear that
$2\omega_0$ has to be identified with $\mu_{II}$, a scale in SCET$_{\rm II}$.
 It is  thus not to be  expected that (\ref{eq:1})
exactly reproduces the logarithms resummed in the
Wilson-coefficients $C_i$, Eq.~(\ref{wilsons}), which are characteristic
for SCET$_{\rm I}$. Indeed, subtracting $C_4(m_b/(2\omega_0)) [1 +
3 a_s \ln(m_b/(2\omega_0))]$ from (\ref{eq:1}) (the second logarithmic
term accounts for the $\mu$-dependence of $f_B\sqrt{m_B}$ in the HQL),
the double logs cancel, but 
a term $-4 \ln m_b/(2\omega_0)$ remains. This term is
in fact universal for all form factors contributions, including those
listed in the appendix. This suggests that they are related to matching
effects of 
SCET$_{\rm II}$ onto SCET$_{\rm I}$ which are not included in
(\ref{eq:1}). 

\begin{figure}
$$\epsffile{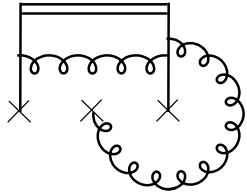}$$
\caption[]{$O(\alpha_s)$ contribution of ${\cal T}$ to $f_+$. Double
  line: b quark propagator, lines ending with a cross: onshell
  partons.}\label{fig1}
\end{figure}

As for the contributions to $f_+$ induced by the twist-3 DAs $\phi_p$
and $\phi_\sigma$, at first glance they do not appear to follow the
pattern displayed by $f_+^{T2}$: there are mixed logarithms
$\sim \ln (2\omega_0/\mu) \ln (2\omega_0/m_b)$. These logarithms
 disappear upon implementing the equation of motion constraint
$\phi'_\sigma(1,\mu) = - 6 \phi_p(1,\mu)$, which follows
from the conformal expansion of the DAs discussed in
Ref.~\cite{update} and is valid exactly (i.e.\ to all orders in the
conformal expansion). There is a similar relation
for $u=0$:  $6\phi_p(0,\mu) = \phi'_\sigma(0,\mu)$. One then has
\begin{eqnarray}
f_B\, m_b\, f_+^{T3}(0)  &= & f_B\, m_b ( f_+^{T3,p}(0) +
f_+^{T3,\sigma}(0)) \nonumber\\
&=&  2\mu_\pi^2(\mu)\,\frac{\omega_0}{m_b}\,\phi_p(1,\mu) 
\left\{ 1 + a_s \left( \pi^2 - 7
-2\ln^2\,\frac{2\omega_0}{m_b} -4\ln\,\frac{2\omega_0}{m_b} + 6
\ln\, \frac{2\omega_0}{\mu}\right)\right\}\nonumber\\
&&{} + 2\mu_\pi^2(\mu)\,\frac{\omega_0}{m_b}\, a_s\,
\left(4\ln\frac{2\omega_0}{ \mu}-3\right) \int_0^1 du\,\frac{\phi_p(u)
  -\phi_p(1)}{\ub} \nonumber\\
&&{}- \mu_\pi^2(\mu)\,\frac{\omega_0}{3m_b}\, a_s\,\int_0^1 du\,\frac{
\phi_\sigma(u) + \ub\phi'_\sigma(1)}{\ub^2}\,. \label{boo}
\end{eqnarray}
The  term in $6\ln (2\omega_0/\mu)$ cancels the $\mu$-dependence of 
$\mu_\pi^2$. As for the remaining logarithms in
$2\omega_0/\mu$, they contribute only at nonleading order in the
conformal expansion of $\phi_p$ and do not cancel the $\mu$-dependence
of $\phi_p(1,\mu)$. This is to be expected as, as discussed in Sec.~2,
$\phi_p$ mixes with the 3-particle DA ${\cal T}$ and we expect
only the combination $f_Bm_b (f_+^{T3,p} + f_+^{T3,\sigma} +
f_+^{T3,{\cal T}})$ to be $\mu$-independent. At first glance,
this appears to be in contradiction with the HQL of the tree-level
contribution of ${\cal T}$, which is suppressed by one power of
$m_b$, which indeed has caused some confusion, cf.~\cite{Lange2}. The
power-suppression of the tree-diagram is however a consequence of the fact
that the gluon line is emitted from the b quark and is expected to be
removed at $O(\alpha_s)$, as  the complete expression
$f_+^{T3}$ must be $\mu$-independent.
To verify this expectation, I have calculated the
diagram shown in Fig.~\ref{fig1}. 
In the HQL and in Feynman gauge, 
it yields (using the asymptotic form of ${\cal T}$)
\begin{equation}
f_B m_b f_+^{\mbox{\scriptsize 
Fig.\ref{fig1}}}(0) = 2 \mu_\pi^2 \,\frac{\omega_0}{m_b}\,
a_s (-160 \eta_3),
\end{equation}
i.e.\ the power-suppression effective at tree-level is indeed removed.
Numerically, $\eta_3(1\,\mbox{GeV})$ $\approx 0.01$ \cite{update}. The
reason for the huge factor 160 is the factor 360 in the asymptotic
DA, Eq.~(\ref{asymptotic}). The size of this contribution has to be
compared with the one induced by the integral over $\phi_p$ in
(\ref{boo}), which gives 270, in the same units. Assuming that the
diagram in Fig.~\ref{fig1} yields a typical contribution, the
$O(\alpha_s)$ corrections to ${\cal T}$ should turn out
to be of roughly the same size as the $O(\alpha_s)$ contributions
induced by $\phi_p$ and $\phi_\sigma$, which, in full LCSRs
without the finite-energy limit $M^2\to\infty$ are nonnegligible, but
not sizeable \cite{roman}. 

Using the formulas collected in App.~\ref{app:A}, it 
is straightforward to verify that LCSRs also reproduce the
structure of symmetry-breaking hard vertex-corrections
\begin{eqnarray}
f_+(0) + f_-(0) & = & a_s \xi(E_{\rm max}) + \dots \nn\\
f_T(0,\mu) - f_+(0) & = & a_s \xi(E_{\rm max})\left( 2 +
\ln\frac{m_b^2}{\mu^2} \right) +\dots
\end{eqnarray}
with 
$$\xi(E_{\rm max}) = \frac{1}{f_Bm_b}\left(
-f_\pi\,\frac{\omega_0^2}{m_b}\, \phi'_\pi(1) + 2 \mu_\pi^2\,
\frac{\omega_0}{m_b}\,\phi_p(1) + O(\alpha_s)\right),
$$
where the dots denote integral terms $\sim\int_0^1 du$.

The last term in (\ref{eq22}) and (\ref{eq:blurb}) that has to be
identified with a corresponding structure in the LCSRs is $\Delta F$,
Eq.~(\ref{eq:deltaf}). In order to do so, one has to express  
$1/\lambda_B$
in terms of a finite-energy sum rule. Such a sum rule has been obtained
in Ref.~\cite{emi} from a comparison of the SCET factorised expression
for the $B\to\gamma$ form factor and its SVZ sum rule and reads:
\begin{equation}\label{eq:lambda}
f_B^2 m_b\,\frac{1}{\lambda_B} = \frac{3}{2\pi^2} \,\omega_0^2.
\end{equation}
$\Delta F$ should thus translate into 
$$
f_B m_b\Delta F = 4 f_\pi\,\frac{\omega_0^2}{m_b}\,\int_0^1
du\,\frac{\phi_\pi(u)}{\ub}.
$$
Indeed, I find
\begin{eqnarray}
f_B m_b(f_T^{T2}(0) - f_+^{T2}(0)) & = & 
-4 f_\pi\,a_s\,\frac{\omega_0^2}{m_b}\,\int_0^1
du\,\frac{\phi_\pi(u)}{\ub} + \mbox{terms with no integral},\nonumber\\
f_B m_b(f_T^{T3}(0) - f_+^{T3}(0)) & = & 0 + \mbox{terms with no
  integral},\nonumber\\
f_B m_b(f_+^{T2}(0) + f_-^{T2}(0)) & = & 
4 f_\pi\,a_s\,\frac{\omega_0^2}{m_b}\,\int_0^1
du\,\frac{\phi_\pi(u)}{\ub} + \mbox{terms with no
  integral},\nonumber\\
f_B m_b(f_+^{T3}(0) + f_-^{T3}(0)) & = & 0 + \mbox{terms with no
  integral}.
\end{eqnarray}
This pattern of integral terms coincides with the SCET results
(\ref{eq22}) and (\ref{eq:blurb}) and implies that to leading order in $1/m_b$ 
symmetry-breaking hard spectator interactions indeed involve only the
twist-2 DA $\phi_\pi$. Although an  explicit calculation of
$O(\alpha_s)$ contributions to ${\cal T}$ is not available, we do
not expect them to break spin-symmetry: the three twist-3
DAs $\phi_p$, $\phi_\sigma$ and ${\cal T}$ are coupled by evolution
equations, so the presence of symmetry-breaking contributions from ${\cal T}$ 
would entail corresponding contributions from $\phi_p$ and $\phi_\sigma$. The
only way this could be avoided is if ${\cal T}$ observes a evolution
equation of its own, i.e.\ if it mixes into $\phi_p$ and
$\phi_\sigma$, but not vice versa.

So far we have essentially verified that LCSRs reproduce
the relations for the differences
of form factors derived in Ref.~\cite{BF}. The next question to ask is
if the symmetry-breaking factorisable term in 
(\ref{eq:SCETfuck}) can also be derived
from LCSRs. To this end, we look more closely at the calculation of
radiative corrections to the correlation function (\ref{eq5}).
The one-loop Feynman diagrams are
shown in Fig.~\ref{fig2}. It turns out that to twist-2 accuracy 
they can be expressed in terms of a few basic traces:
\begin{figure}
$$\epsfxsize=10cm\epsffile{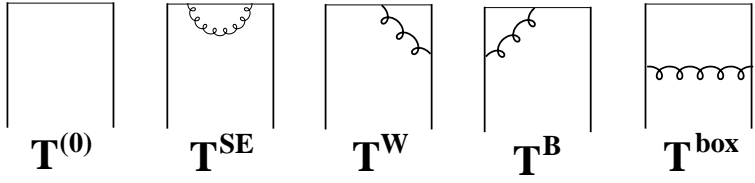}$$
\caption[]{Radiative corrections to the correlation function
  (\ref{eq5}). The external quark lines are onshell with momenta $up$
  and $\ub p$, respectively. $T^B$, the radiative correction to the B
  vertex, and $T^{\rm box}$ correspond to hard
  spectator interaction diagrams in SCET.}\label{fig2}
\end{figure}
\begin{eqnarray}
Tr_1 & = & Tr\left(\pro\Gamma\slash{q}\gp\right) \equiv 
Tr\left(\pro\Gamma\slp_B\gp\right),\nonumber\\
Tr_2 & = & Tr\left(\pro\Gamma\gp\right),\nonumber\\
Tr_3 & = & Tr\left(\pro\slash{q}\Gamma\gp\right) \equiv
Tr\left(\pro\slp_B\Gamma\gp\right),\nonumber\\
Tr_4 & = & Tr(\pro\slash{q}\Gamma\slash{q}\gp). 
\end{eqnarray}
${\cal P}$ is the projector onto the final state meson, ${\cal P}_\pi
= \slash{p}\gamma_5$, and $\Gamma$ is the Dirac structure of the weak
vertex. The contribution of each diagram to the correlation function 
is given by
$$\frac{i}{4}\,f_\pi\int_0^1 du\,\phi_\pi(u) T(u; p_B^2,q^2).$$
The tree-level diagram yields
$$T^{(0)} = \frac{i}{s}\,(Tr_1 + m_b Tr_2),$$
with $s = m_b^2 - u p_B^2 -\ub q^2$.
The analogues of the hard spectators diagrams are $T^B$ and $T^{\rm
  box}$; they are given by
\begin{eqnarray}
T^B & = & 2\,\frac{g^2C_F}{s}\left\{ \left(-8\bar
C(1+\epsilon)-1-m_b^2\bar B + \ub (p_B^2-q^2) \bar A\right) 
(Tr_1+m_b Tr_2) -m_b s\bar B Tr_2\right\},\nn\\
T^{\rm box} & = & -g^2 C_F a_\pro \left\{ a_\pro H (Tr_1 + m_b Tr_2) +
I (-m_b Tr_4 + (s-m_b^2) Tr_3) + \widetilde{B} Tr_3\right\}\label{Ts}
\end{eqnarray}
with
$$
a_\pro \pro := \gu \pro \go \mbox{~~and~~} \widetilde{B} = B (u\to 1).
$$
The integrals are defined as
\bea
\intk {k_\al \over (k+up)^2 k^2 [(q-k)^2-m^2]} &=& A up_{\al}+Bq_\al,\nn \\
\intk {k_\al \over (k-\ub p)^2 k^2 [(p_B+k)^2-m^2]} &=& -\bar A \ub 
p_{\al}- \bar B p_{B\al},\nn \\
\intk {k_\al  k_\be  \over k^2  (k-u p)^2 (k+\bx p)^2
[(u p+q-k)^2-m^2]} &=  & H g_{\al\be}+I q_\al
q_\be +\cdots,\nn
\eea
where the dots stand for irrelevant terms. All infrared divergent
terms are treated in dimensional regularisation and can be absorbed
into the distribution amplitude. All occurring convolution integrals
are finite; there are no endpoint singularities.

The terms in $Tr_1 + m_b Tr_2$ are obviously spin-symmetric; for
$f_+(0)$, one has $Tr_1=Tr_3=Tr_4=0$ and $Tr_2=-2$. The only
spin-symmetry breaking contribution to the correlation function
$\Pi_+$ is then
\begin{equation}
\Pi_+^{\mbox{\scriptsize sym-break}} = i f_\pi a_s m_b (4\pi)^2 \int_0^1 du\,
\phi_\pi(u) \bar B(u,p_B^2).
\end{equation}
As shown in App.~\ref{app:B}, this expression yields
\begin{equation}
f_B m_b f_+^{\mbox{\scriptsize sym-break}}(0) = 2 a_s f_\pi
\,\frac{\omega_0^2}{m_b}\, \int_0^1 du\,\frac{\phi_\pi(u)}{\ub} = f_B
m_b \left( \frac{4\pi^2}{3}\, a_s \,\frac{f_Bf_\pi}{m_b}\,
\frac{1}{\lambda_B}\, \left\langle
\frac{1}{\ub}\right\rangle_\pi\right),
\end{equation}
which conincides with the r.h.s.\ of (\ref{eq:SCETfuck}) for
$E_\pi=m_B/2$ and to $O(\alpha_s)$ accuracy.\footnote{Up to the replacement
$u\leftrightarrow \ub$ under the integral, which is justified for
$\phi_\pi$, but  not for $\phi_K$.}

The last point on the list of items on p.~\pageref{roadmap} still 
to be investigated is the possible
emergence of symmetry-breaking nonfactorisable terms at nonleading
order in the heavy quark expansion. Such terms have been found in
an expansion of the SCET Lagrangian to second order \cite{sup}, and we
find them also in the sum rule approach. Restricting ourselves to
tree-level, we have
\begin{equation}
f_B m_B (f_T^{T3}(0) - f_+^{T3}(0)) = -4 \mu_\pi^2
\,\frac{\omega_0^2}{m_b^2} \phi_p(1).
\end{equation}
This evidently soft symmetry-breaking terms confirms the conclusion of
\cite{sup} that simple sym\-me\-try-\-re\-la\-tions for form factors are broken
by nonfactorisable power-suppressed corrections. We will discuss the
numerical impact of these terms in the next section.

To summarise, I have found that, to $O(\alpha_s)$ and twist-3
accuracy, and in the heavy quark limit,
\begin{itemize}
\item LCSRs for the form factors of $B\to\pi$ transitions observe the
  relations for spin-symmetry breaking corrections obtained in Ref.~\cite{BF};
\item LCSRs predict the same symmetry-breaking hard-scattering
  amplitude as the SCET factorisation theorem \cite{fuck};
\item at $E_\pi = m_b/2$, LCSRs reproduce the structure of double
  logarithms in $m_b$, but
  not of single logs, which I attribute to yet unknown matching
  effects from SCET$_{\rm II}$ onto SCET$_{\rm I}$;
\item the $1/m_b$ suppression of 3-particle twist-3 contributions at
  $O(1)$ is removed at $O(\alpha_s)$, as required by
  the $\mu$-independence of $f_+$; as all effects of 2-particle twist-3
  DAs are symmetry-preserving to $O(\alpha_s)$, and 2 and 3-particle
  DAs are linked by evolution equations, the same is likely to
  happen also
  for the 3-particle ones, so that it is expected that 
  all twist-3 effects can be absorbed into $\zeta$;
\item LCSRs are free of endpoint-singularities throughout, including
  the sample diagram Fig.~\ref{fig1}; it should
  be possible to prove that within SCET itself, along precisely the
  same lines as other factorisation proofs.
\end{itemize}
At the end of this lengthy section, let me also state clearly where
LCSRs fall behind SCET factorised expressions: the resummation of
large logarithms appears difficult, in particular since in the
finite-energy limit there is no obvious candidate for the intermediate
scale $\sim \sqrt{m_b\Lambda_{\rm QCD}}$. It is possible that in full
sum rules the Borel parameter may play that part, as according to
(\ref{scaling}) it has the correct scaling in $m_b$. Also, the interpolation
of the B meson by a local current entails certain
differences between the two approaches: 
the B meson DA $\phi_B^+$ does not enter LCSRs explicitly
and in order to verify the structure of the factorisable term
it was necessary to invoke a second sum rule for $1/\lambda_B$. It
is very likely that any finer details relating to the interactions of the
spectator quark in the B, for instance the description of how the
soft spectator turns into a collinear one in the
$\pi$, i.e.\ any information about the jet functions $J$ in
(\ref{eq:SCETfuck}), cannot be resolved in the LCSR approach. Rather
than considering that a weakness, I would like to stress
that the intention of LCSRs is not so much to study the intricacies of
QCD in the heavy quark \& large energy limit, 
and in that respect they certainly cannot compete with
SCET, but to provide phenomenologically relevant 
numerical predictions within a well-defined and controlled framework
and in full QCD. 

\section{Numerics}

After the more formal discussion in the last section which has
demonstrated that LCSRs in the heavy quark limit fulfill all SCET
relations and constraints, we proceed to the analysis of numerical
aspects.

The numerical value of $f_+(0)$ is comparatively well known, from both
sum rules and lattice calculations; both methods point
consistently at the same result, $f^{\rm QCD}_+(0)
\approx 0.3$ \cite{roman,xx}. Let us
compare this number with the contribution of the factorisable term in
Eq.~(\ref{eq:SCETfuck}):
$$f_+^{\mbox{\scriptsize fac}}(0) =
\frac{4\pi^2}{3}\,a_s\,
\frac{1}{\lambda_B}\, \left\langle \frac{1}{u}\right\rangle_{\pi}.
$$
The first inverse moment of the twist-2 $\pi$ DA can be extracted from
experiment \cite{cleo}, at the scale $\mu\approx 1\,$GeV:
$$
\left\langle \frac{1}{u}\right\rangle_{\pi} = 3.3\pm 0.3.
$$
As for the inverse moment of the B meson DA, $1/\lambda_B$, it has
recently be determined from QCD sum rules as $\lambda_B = 0.6\,$GeV
\cite{emi}. As this determination corresponds to a tree-level sum
rule, the numerical value has to be taken cum grano salis, but should
be sufficient for a rough estimate of the factorisable term.
$\alpha_s$ is to be evaluated at the intermediate 
scale $\mu^2\sim \Lambda_{\mbox{\scriptsize QCD}} m_b$. 
To enhance the term as much as possible, I
choose $\mu=1\,$GeV, i.e.\, $\alpha_s=0.51$. Taking all together,
this yields
\begin{equation}
f_+^{\mbox{\scriptsize fac}}(0) \approx 0.023,
\end{equation}
i.e.\ about 10\% of the total form factor $f_+^{\mbox{\scriptsize
    QCD}}(0) \approx 0.3 $.

To compare the size of the factorisable term with that 
of power-suppressed corrections, the obvious thing
to do is to derive a sum rule for $\xi$ (or $\zeta$) and compare
with the full QCD result. To NLO in the conformal 
expansion of DAs I find:
\begin{eqnarray}
\lefteqn{f_B m_b^2 \xi(E_{\rm max}) e^{-\bar\Lambda/\tau} =}\nn\\
& = & 12f_\pi (1+6
a_2^\pi(\mu)) \int_0^{\omega_0} d\omega  \omega  e^{-\omega/\tau}\left\{ 1 +
a_s \left( \pi^2+ 3 - 2 \ln^2 \frac{2\omega}{m_b} - 6
\ln\frac{2\omega}{m_b} \right)\right\}\nn\\
&&{} + 72 f_\pi a_2^\pi(\mu)
\int_0^{\omega_0} d\omega  \omega  e^{-\omega/\tau} a_s \left(
\frac{5}{2} - \frac{25}{3}\,\ln\frac{2\omega}{\mu}\right)\nn\\
&&{} + 2\mu_\pi^2(\mu)
(1+30\eta_3(\mu) - 3 \eta_3(\mu) \omega_3(\mu)) \int_0^{\omega_0}
d\omega e^{-\omega/\tau}\nn\\
&&{} + 2\mu_\pi^2(\mu) a_s \int_0^{\omega_0}
d\omega e^{-\omega/\tau} \left(\pi^2 - 4 - 2 \ln^2 \frac{2\omega}{m_b}
- 8 \ln\frac{2\omega}{m_b} + 6 \ln\frac{2\omega}{\mu}\right).\label{shitxi}
\end{eqnarray}
In the finite-energy limit $\tau\to\infty$, this agrees with
(\ref{eq:1}). 
One point to notice is that, although the authors of \cite{SCET,fuck} 
have argued that the resummation of
single and double logarithms in $\ln m_b/\mu_F$ be relevant,
this is not obvious from a phenomenological point of view:
$\sim\ln^2 m_b/(2\omega_0)\sim \ln^2 (5/2) \approx 1$ is not really large
and in fact much smaller than the nonlogarithmic contributions.

Unfortunately, it turns out that the above sum rule yields $\xi
\approx 2 $ in complete disagreement with 
the expected result $\sim 0.3$. The reason
is that the expansion of the sum rule for $f_+(0)$ in  
$1/m_b$ is extremely bad at $m_b=5\,$GeV.
\begin{figure}
$$\epsfxsize=0.53\textwidth\epsffile{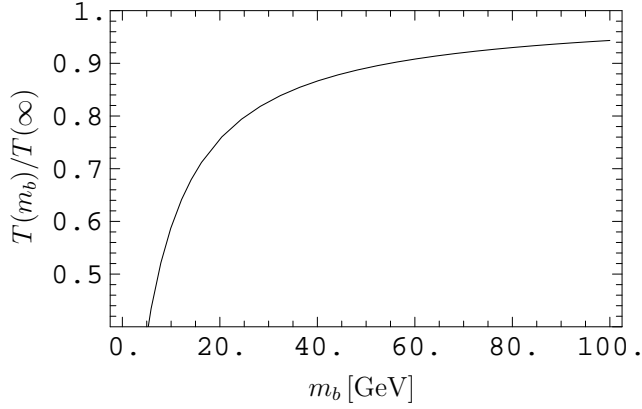}$$
\vskip-0.5cm
\caption[]{Tree-level twist-2 contribution to $f_+(0)$ as a function
  of $m_b$. Input parameters: $\bar\Lambda = 0.5\,$GeV, $a_2 = 0$,
  $\omega_0=1\,$GeV, $\tau = 0.7\,$GeV.}\label{fig:mb}
\end{figure}
Fig.~\ref{fig:mb} illustrates this feature: I plot the dependence of
the tree-level twist-2
 contribution to $f_+(0)$ on $m_b$, 
\begin{eqnarray}
T(m_b) &\equiv & 
f_B m_b^2 f_+^{T2,{\rm QCD}}(0)\nn\\
& = & - 6 f_\pi \int_0^{\omega_0} d\omega
\exp\left(\frac{\bar\Lambda -\omega}{\tau}\left[ 1 + \frac{\bar\Lambda +
    \omega}{2m_b} \right]\right) \frac{m_b^4}{(m_b+\omega)^3} \left(
\frac{m_b^2}{(m_b+\omega)^2} - 1\right) \nn\\
&&\times\left( 1 + 6a_2 \left(1 - \frac{5
  m_b^2 \omega (2 m_b+\omega)}{(m_b+\omega)^4}\right)\right)
\end{eqnarray}
relative to its limiting value for $m_b\to\infty$, as given in the
 first term on the r.h.s.\ of
 Eq.~(\ref{shitxi}). It is obvious that  Eq.~(\ref{shitxi}) overshoots
 the value at the physical quark mass by more than a factor 2. 
As the calculation of $\xi$ has, to the best of my knowledge,
never been attempted by any other method, it is not clear if 
Eq.~(\ref{shitxi}) indeed indicates that power-corrections to $\xi$ are
 huge, with the asymptotic value only reached at quark masses
$m_b\sim 100\,$GeV. A calculation of $\xi$ in the
pure heavy quark limit from lattice, if possible, would surely help to
clarify the situation.

\begin{figure}
$$\epsfxsize=0.48\textwidth\epsffile{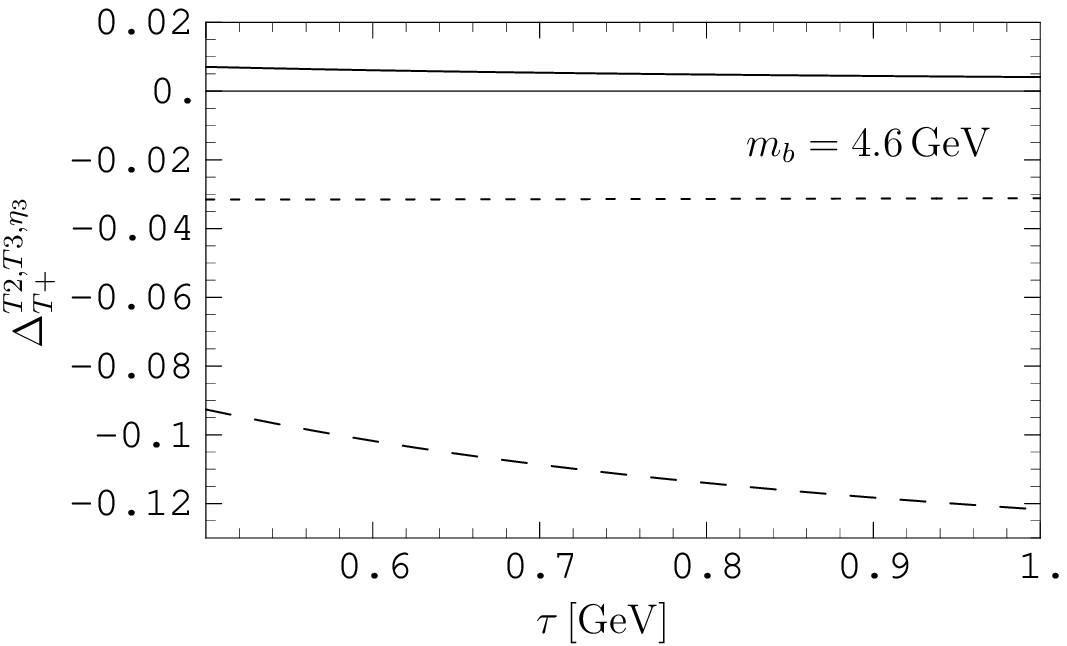}\quad
\epsfxsize=0.48\textwidth\epsffile{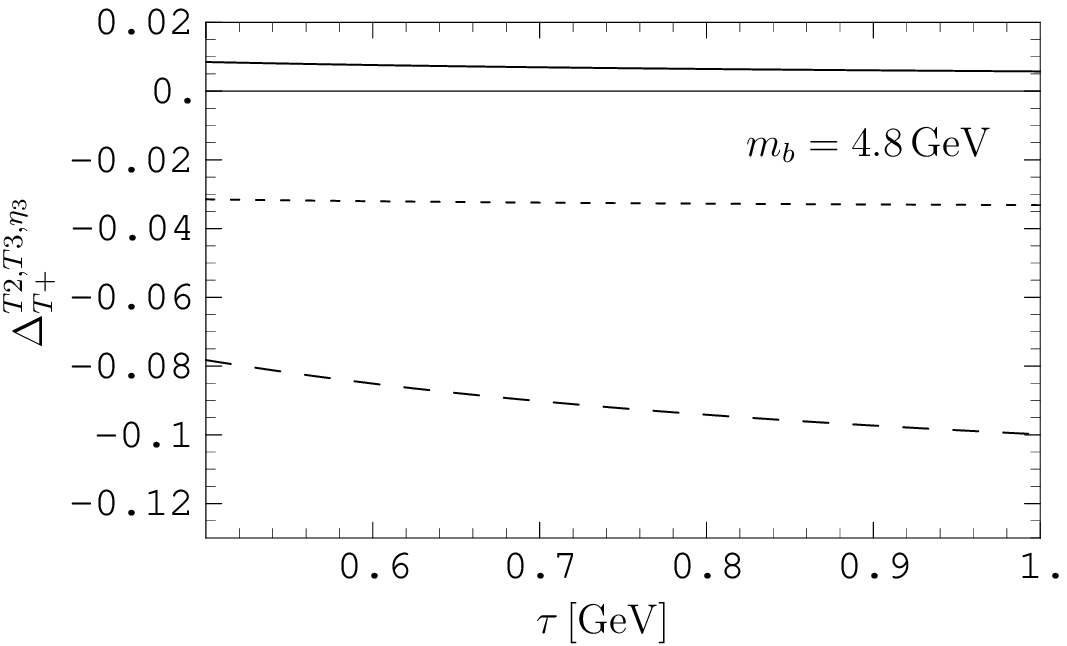}$$
\vskip-0.5cm
\caption[]{Relative differences $\Delta_{+T}$ between $f_T(0)$ and
  $f_+(0)$ induced by twist-2 and 3 contributions, respectively, for
  two different values of the b quark mass. Solid line: radiative
  corrections to the twist-2 contribution; long dashes: 2-particle
  twist-2 DAs (asymptotic form), including $O(\alpha_s)$ corrections; 
  short dashes: contribution of
  $\eta_3$, from both 2 and 3-particle DAs (tree-level).}\label{fig4}
\end{figure}
As an alternative, one can calculate the difference $f_T(0)-f_+(0)$
directly from LCSRs, which feature an {\em
  exact} (i.e.\ all-order in $1/m_b$)
cancellation of the tree-level twist-2 contribution, so the difference
should be sensitive to symmetry-breaking effects induced at
higher-twist. Let me define (choosing $\mu=m_b$ for the ultraviolet
scale in $f_T$)
\begin{equation}
\Delta_{T+}^{(i)} = \frac{f_T^{(i)}(0) - f_+^{(i)}(0)}{f_+^{\rm QCD}(0)}
\end{equation}
with $i\in\{T2, T3, \eta_3,T4\}$, i.e.\ the relative contributions to
  the form factor difference from twist-2, 2-particle twist-3,
  3-particle twist-3 and twist-4 contributions, respectively. 
All these terms are
  calculated in full QCD, with no $1/m_b$ expansion; for $f_T$, the
  radiative corrections to the twist-3 contributions are new. The input in the
  corresponding sum rules is the same as in Ref.~\cite{roman}: the b
  quark mass (one-loop pole mass) is fixed at 4.6 (4.8)~GeV. The
  continuum threshold $\omega_0 = 1.25\,$GeV, i.e.\ $s_0 = 34\,{\rm
  GeV}^2$, together with the Borel-parameter window
$0.5\,{\rm GeV}< \tau < 1\,$GeV yields $f_B =
  0.195\,$GeV, in good agreement with lattice results. I fix $\mu =
  \sqrt{m_B^2-m_b^2}$, i.e.\ it is formally of order $\sqrt{m_b}$.

The results are shown in Fig.~\ref{fig4}. $\Delta_{T+}^{T2}$ is
  formally of leading order in $1/m_b$, but $O(\alpha_s)$, and
  actually is very small numerically, $\sim\,$1\%. This term includes
  in particular the hard symmetry-breaking terms in (\ref{eq:blurb}),
  which  numerically cancel to a large extent. The 2-particle twist-3
  term is formally $O(1/m_b)$, but starts at tree-level. It is
  numerically the most dominant effect, a nonfactorisable
  symmetry-breaking term that  cannot be calculated in SCET (cf.\ also
  Ref.~\cite{sup}). The (tree-level) contribution of the
  quark-quark gluon matrix element $\eta_3$ that enters both
  nonleading terms in the conformal expansion of $\phi_p$ and
  $\phi_\sigma$ and the leading term in ${\cal T}$ is shown
  separately, as $\Delta_{T+}^{\eta_3}$; it is small and corrects
  the form factor by about 3\%. Finally, and not shown in the plot,
  the contribution of twist-4 DAs is of about the
  same size as the factorisable twist-2 ones: 1\%. The most
  important contribution still missing are radiative corrections to
  the 3-particle twist-3 DA, which are formally of leading order in
  $1/m_b$ for $f_T$ and $f_+$ separately, 
  but expected to be spin-symmetric, i.e.\ of order $\alpha_s/m_b$ in
  the difference of form factors.  

A similar analysis could be performed for $\Delta_{-+}$, but even
without that the overall emerging picture is
quite clear: nonfactorisable symmetry-breaking corrections are the
numerically most relevant ones, whereas factorisable contributions are
small. 

The above results should not be interpreted as a new determination of
$f_T(0)$; updates for all form factors will be published separately.

\section{Summary and Conclusions}

In this paper I have discussed, in quite some detail, the relation between SCET
factorisation formulas for the $B\to\pi$ form factors $f_{+,0,T}$ and
their light-cone sum rules. I have demonstrated that to $O(\alpha_s)$
accuracy and for contributions from 2-particle twist-2 and 3 distribution
amplitudes, light-cone sum rules in the large energy limit
$m_b\to\infty$, $E_\pi\to\infty$ fulfill all symmetry relations
predicted by SCET. I have also demonstrated that the nonfactorisable
contributions from 2-particle twist-3  DAs are spin-symmetric to
leading order in $1/m_b$, but break spin-symmetry at higher order.
Numerically they induce a splitting between $f_+(0)$ and $f_T(0)$ of
about 10\% and are the most relevant symmetry-breaking
corrections.\footnote{This is not a statement about the actual
  splitting between these two form factors, but about its
  contribution from 2-particle twist-3 DAs. A more comprehensive study
  is underway.} Within the SCET approach, these contributions correspond
to subleading soft form factors and cannot be resolved any further,
cf.\  Ref.~\cite{sup}. For the chosen form factor difference
$f_T-f_0$, the hard gluonic corrections cancel to a large extent; this
need not be the case for other form factor differences, where these
corrections may even compete with the subleading ones. The results of
my study suggest, however, that subleading soft form factors are, in general,
likely to be numerically relevant in form factor relations, also for
$B\to$ vector meson decays.

The motivation for this study was the question if SCET predictions 
can help to get a better grip on B decays. There is no doubt that SCET
constitutes a major step forward and 
has transformed the way factorisation proofs will be done, with its
shift from a diagrammatical level to all-order proofs based
on field transformations in the Lagrangian. However, in the context of
for instance electromagnetic hadron form factors, experience has shown 
that factorisation is often broken at subleading level, i.e.\ that the
clean separation between hard and soft contributions is spoiled at the
first nontrivial order in the expansion. Usually this is not a problem from a 
phenomenological point of view as, if the momentum transfer is large
enough, experiments are insensitive to these power-suppressed
effects. The situation is different in B decays: the expansion
parameter, $1/m_b$, is not a dynamical variable, but 
fixed and not very small. Experience with HQET, the heavy quark
effective theory, has shown that $1/m_b$ corrections in general
cannot be neglected, and one of the most important applications of
HQET in B physics, the extraction of $|V_{cb}|$ from $B\to D^* e\nu$
at zero recoil, relies precisely on the fact that linear corrections
in $1/m_b$
are absent, by virtue of Luke's theorem.
In the application of SCET to heavy-to-light decays, nonfactorisable
terms are present already at leading order\footnote{The reason being that,
  technically speaking, the operators entering the factorisable
  part in (\ref{eq:SCETfuck}) are subleading in the expansion
  parameter of the SCET Lagrangian, so that it would be more precise
  to say: factorisable contributions are {\em absent} at leading
  order.} and make it impossible to
calculate for instance $f_+$ directly. Still, one may expect 
to express the ratio of form factors in terms of perturbatively calculable
coefficients. The results I have presented in this paper indicate that
this expectation is not realised. The same is likely to be the case also for 
other B decay processes treated in a similar
way, in particular nonleptonic decays. 

Let me conclude with a comment on the prospects of relating $B\to\pi$ to
$B\to K$ form factors via SU(3) symmetry. As $\zeta(E_{\rm max})$ 
is essentially proportional to $-f_\pi \phi'_{\pi}(1)$, one expects an
SU(3) breaking ratio
\begin{equation}\label{xi}
\frac{\zeta_K(E_{\rm max})}{\zeta_{\pi}(E_{\rm max})} \equiv
1 + \Delta_{\rm SU(3)}^\zeta(\mu) \approx 
\frac{f_K\phi'_K(1,\mu)}{f_\pi \phi'_{\pi}(1,\mu)} =
\frac{f_K}{f_\pi} \,\frac{2+\sum_{n=1}^\infty a_n^{K}(\mu) (n+1)(n+2)}{2+
\sum_{n=1}^\infty a_n^{\pi}(\mu) (n+1)(n+2)},
\end{equation}
where we have expressed $\phi(u)$ by its conformal expansion
$$\phi(u,\mu) = 6 u (1-u) (1 + \sum_{n=1}^\infty a_n(\mu) C_n^{3/2}(2u-1)).$$
The first two Gegenbauer moments $a_1^K$ and $a_2^K$ have recently be
redetermined  in Ref.~\cite{BB03}:
$$a_1^K(1\,{\rm GeV}) = -0.18\pm 0.09, \quad a_2^K(1\,{\rm GeV}) =
0.16\pm 0.10.$$
For the $\pi$, there are experimental data available as well as sum
rule calculations. Taking the number
quoted in a recent paper, Ref.~\cite{Bakulev}, $a_2^{\pi}(1\,{\rm GeV})
= 0.19$ ($a_1^\pi=0$ up to isospin-breaking effects), one has
$$\Delta_{\rm SU(3)}^\zeta(1\,{\rm GeV}) = -0.18,$$
which indicates nonnegligible SU(3) breaking.
Although this number comes with a large uncertainty, as it is
very sensitive to higher order
Gegenbauer moments, it indicates that the actual SU(3) breaking can be
large and potentially even have different sign than the naive expectation
$\Delta_{\rm SU(3)}^\zeta\approx f_K/f_\pi-1 = 0.2$.

\section*{Acknowledgements}
I would like to 
thank the participants of the Ringberg Workshop on Heavy Flavours,
Ringberg April 2003, and FPCP, Paris June 2003, for stimulating
presentations and discussions, which contributed significantly to the
genesis of the considerations presented in this paper.

\appendix

\setcounter{equation}{0}
\renewcommand{\theequation}{A.\arabic{equation}}

\section{Light-Cone Sum Rules in the Heavy Quark Limit}
\label{app:A}

In this appendix I give the explicit LCSRs for $f_-(0)$ and $f_T(0)$
in the combined heavy quark and finite energy limit, for 2-particle
twist-2 and 3 contributions to $O(\alpha_s)$ accuracy. 
Use of the equation of motion
relation $6\phi_p(1) = -\phi'_\sigma(1)$ is implied. 
I use the notations  $\mu_\pi^2 = f_\pi
m_\pi^2/(m_d+m_u) = -2\langle \bar q q \rangle/f_\pi$ and $a_s = C_F
\alpha_s/(4\pi)$.
\begin{eqnarray}
\lefteqn{f_B\, m_b\, f_-^{T2}(0)  = f_\pi\,\frac{\omega_0^2}{m_b}\,
 \phi_\pi'(1) \left\{ 1 + a_s \left( \pi^2
-2\ln^2\,\frac{2\omega_0}{m_b} -4\ln\,\frac{2\omega_0}{m_b} + 2
\ln\, \frac{2\omega_0}{\mu}\right)\right\}}\hspace{2cm}\nonumber\\
&&\hspace{-1cm} - 4 a_s\,f_\pi\,\frac{\omega_0^2}{m_b}\left(
\ln\,\frac{2\omega_0}{ \mu}-1\right) \left\{ \int_0^1 du\,\frac{\phi_\pi(u)
  +\bar u \phi_\pi'(1)}{\bar u^2} + \int_0^1
du\,\frac{\phi_\pi(u)}{\bar u}\right\},\\
\lefteqn{f_B\, m_b\, f_-^{T3}(0)  = -2 \mu_\pi^2
  \frac{\omega_0}{m_b}\, \phi_p(1) \left\{ 1 + a_s \left(\pi^2-8 - 
2 \ln^2 \,\frac{2\omega_0}{m_b}
-4\ln\,\frac{2\omega_0}{m_b}  + 6 \ln\,
\frac{2\omega_0}{\mu}\right)\right\}
}\hspace*{2cm}\nonumber\\
&&{}-2 \mu_\pi^2 \frac{\omega_0}{m_b}\, a_s \left(
4\ln\,\frac{2\omega_0}{\mu} - 3 \right) \int_0^1 du\,\frac{1}{\ub}\,
(\phi_p(u) - \phi_p(1))\nonumber\\
&&{} + \mu_\pi^2\,\frac{\omega_0}{3m_b}\, a_s
\int_0^1 du\,\frac{1}{\ub^2}\,
(\phi_\sigma(u) + \ub \phi_\sigma'(1)),\\
\lefteqn{
f_B\, m_b\, f_T^{T2}(0)  =  -f_\pi\,\frac{\omega_0^2}{m_b}\, 
\phi_\pi'(1) \left\{ 1 + a_s \left( 3+\pi^2
-2\ln^2\,\frac{2\omega_0}{m_b} -6\ln\,\frac{2\omega_0}{m_b} + 4
\ln\, \frac{2\omega_0}{\mu}\right)\right\}}\hspace{2cm}\nonumber\\
&&\hspace{-1cm} + 4 a_s\,f_\pi\,\frac{\omega_0^2}{m_b}\left(
\ln\,\frac{2\omega_0}{ \mu}-1\right) \left\{ \int_0^1 du\,\frac{\phi_\pi(u)
  +\bar u \phi_\pi'(1)}{\bar u^2} + \int_0^1
du\,\frac{\phi_\pi(u)}{\bar u}\right\},\\
\lefteqn{f_B\, m_b\, f_T^{T3}(0)  = 2 \mu_\pi^2
  \frac{\omega_0}{m_b}\, \phi_p(1) \left\{ 1 + a_s \left(\pi^2-5 - 
2 \ln^2 \,\frac{2\omega_0}{m_b}
-6\ln\,\frac{2\omega_0}{m_b}  + 8 \ln\,
\frac{2\omega_0}{\mu}\right)\right\}
}\hspace*{2cm}\nonumber\\
&&{}+2 \mu_\pi^2 \frac{\omega_0}{m_b}\, a_s \left(
4\ln\,\frac{2\omega_0}{\mu} - 3 \right) \int_0^1 du\,\frac{1}{\ub}\,
(\phi_p(u) - \phi_p(1))\nonumber\\
&&{} - \mu_\pi^2\,\frac{\omega_0}{3m_b}\, a_s
\int_0^1 du\,\frac{1}{\ub^2}\,
(\phi_\sigma(u) + \ub \phi_\sigma'(1)).
\end{eqnarray}

\section{\boldmath  Hard spectator contribution to $f_+$}
\label{app:B}
We have
$$\bar B(q^2=0) = \frac{i}{(4\pi^2)}\,
\frac{1}{\ub p_B^2}\left\{ \left( \frac{m_b^2}{u
  p_B^2} -1 \right) \ln\left( 1 - u \,\frac{p_B^2}{m_b^2}\right) -
  \left( \frac{m_b^2}{p_B^2} - 1 \right) \ln \left( 1 -
  \frac{p_B^2}{m_b^2}\right) \right\}.
$$
In the finite-energy limit, the continuum-subtracted Borel transform is
$$
\hat{B}_{\rm sub}^\infty \bar B = \frac{i}{(4\pi)^2}\left[
\frac{1}{\ub}\, (\ln u_0 + \bar u_0)
\Theta (u_0 - u) + \frac{1}{u\ub}\, (u\ln u + \ub u_0)
\Theta (u - u_0)\right]
$$
with $u_0 = m_b^2/s_0$. 
Upon convolution with $\phi_\pi(u)$ and using the scaling law
(\ref{scaling}), the second term yields $8/3\,
\omega_0^3/m_b^3\, \phi'_\pi(1)$, i.e.\ a power-suppressed term, 
whereas the first one yields 
$$
\bar B \to -\frac{i}{(4\pi)^2}\,
\frac{2\omega_0^2}{m_b^2}\,\int_0^1 du\,\frac{\phi_\pi(u)}{\ub}.
$$

\end{document}